\documentclass[aps,prc,twocolumn,superscriptaddress, letter]{revtex4-2}
\newcommand{\sizeA}{1.0}

\usepackage{graphicx}
\usepackage{xcolor}
\usepackage{dcolumn}
\usepackage{amsmath}
\usepackage{wasysym}
\usepackage{siunitx}
\usepackage{comment}

\begin{document}

\title{On Robust $\mathbf{\beta}$-Spectra Shape Parameter Extraction}

\author{B.C. Rasco}
\affiliation{Physics Division, Oak Ridge National Laboratory, Oak Ridge, TN 37831, USA}
\email{rascobc@ornl.gov}
\author{T.J. Gray}
  \affiliation{Department of Physics and Astronomy, University of Tennessee Knoxville, Tennessee 37966, USA}
  \affiliation{Physics Division, Oak Ridge National Laboratory, Oak Ridge, TN 37831, USA}

\author{T. Ruland}
\affiliation{Physics Division, Oak Ridge National Laboratory, Oak Ridge, TN 37831, USA}


\date{\today}

\begin{abstract}
Stable and consistent experimental extraction of $\beta$-shape functions, $C(W)$, is challenging.
Comparing different experimental $\beta$-shapes to each other and to those predicted by theory in a consistent manner is difficult. This difficulty is compounded when different parameterizations of the $\beta$-shape function are used.
Usually some form of a power polynomial of the total electron energy is chosen for this parametrization, but choosing a power polynomial results in extracted coefficients that are highly correlated, with their physical meaning and numerical value dependent on the order of polynomial chosen.
This is true for both theoretical and experimental coefficients, and leads to challenges when comparing coefficients between polynomials of different orders. 
Accurately representing the highly correlated uncertainties is both difficult and subtle and limits the underlying physical interpretation of shape function parameters.
In this manuscript an alternative approach based on orthogonal polynomials is offered.
Orthogonal polynomials offer more stable coefficient extraction with less dependence on the order of the polynomial, allow for easier comparison between theory and experimental coefficients from different order polynomials, can offer reduced correlations between coefficients, while also offering simple physical meaning and clarifying statistical limits of the extracted coefficients.

\end{abstract}

\maketitle

\section{Introduction}
$\beta$-energy spectra have long posed, and even answered some important physics questions.
The $\beta$-energy spectrum originally motivated the identification of the neutrino \cite{pauli30, fermi34} and contributed to identifying the parity non-conservation in $\beta$ decay \cite{wu57, frauenfelder58, beaglehole60}.
Currently, experimental $\beta^{\pm}$-energy spectra are used to predict solar (anti)neutrino energy spectra \cite{bahcall78, winter06, longfellow2023, acharya2025} and reactor antineutrinos \cite{hayes2016, perisse23} at least in the few cases where experimental $\beta$-energy data is available. 
$\beta$-energy spectra may also be used to identify possible beyond standard model physics \cite{gonzalez2019, falkowski21}.
However, if direct data are not available, and this is very often the case, predictions of unmeasured $\beta$-energy spectra are treated as allowed decays.
This may be a poor assumption, especially for forbidden decays \cite{hayes2014, sonzogni_2016}. 
This assumption is often made when evaluating integral electron measurements \cite{wohn73} and in reactor antineutrino summation calculations \cite{fallot12}.
The impact of assuming allowed $\beta$ shapes on total absorption $\beta$-feeding patterns is mentioned in \cite{rasco2016} and discussed in more detail in \cite{shuai22}.
These physics cases are but a few reasons to motivate measuring $\beta$-energy spectra precisely.
However, robust and consistent reporting of $\beta$-energy spectra is limited by the current analysis framework.

There has been much recent experimental interest and activity measuring $\beta$-energy spectra. 
One reason for this renewed interest is due to modern detection techniques, such as magnetic calorimeters \cite{rotzinger08}, superconducting tunnel junction detectors (to measure recoil energy spectra) \cite{fretwell20}, implantation directly into scintillation detectors \cite{hughes19, kanafani22}, along with improved data acquisition systems and simulations to upgrade traditional electron spectral measurements.
In short, it is a very exciting time for $\beta$ decay measurements.

Precise $\beta$-shape predictions are influenced by nuclear structure including factors such as decay energy, angular momentum, nuclear matrix elements \cite{hayen19a, ramalho24, seng25}, and by QED, Electroweak, and other corrections \cite{merzbacher51, hayen18}. 
Being able to compare theoretical and experimental $\beta$ spectra in a clear and consistent manner is essential.
Current $\beta$-shape parameter extraction is usually based on a power-polynomial expansion (with a few modern examples offered here \cite{kossert11, paulsen20, kostensalo23, kostensalo24, craveiro24}). 
This approach has certain challenges. 
Notably, the coefficients extracted using this method depend on the order of polynomial used for the $\beta$-shape function.
This makes meaningful physical interpretation of different coefficients from different experiments and theoretical calculations difficult.
An additional issue with the power-polynomial approach is that the coefficients are highly correlated with each other.
In practice this means different $\beta$-shape coefficients can lead to similar $\beta$-energy spectra.
Without stable and precise (robust) parameters extracting a meaningful physics interpretation for the coefficients is difficult.
This manuscript explores the use of weighted orthogonal polynomials to extract more robust and independent $\beta$-shape coefficients.

Other areas of nuclear physics use orthogonal polynomials to good effect, such as particle-$\gamma$ angular correlations \cite{rose67} and $\gamma$-$\gamma$ angular correlations \cite{bill_mills_2016_45587}. 
Originally, these were calculated using powers of $\cos^{2n}(\theta)$ \cite{hamilton1940}. 
However, formulations which use Legendre polynomials have become more standard.
There are several reasons Legendre polynomials are used.
First is the calculations of $\gamma$-$\gamma$ angular correlation are more straight forward \cite{rose67}.
A second reason to use orthogonal polynomials for $\gamma$-$\gamma$ angular correlations is that it removes the need to change the overall normalization of the coefficients for different powers of $\cos^{2n} (\theta)$ when improved measurements are made or when additional corrections are applied. 
And third, the use of orthogonal polynomials reduces the correlations between the extracted coefficients.
These are all desirable effects that would also be useful in the analysis of experimental and theoretical $\beta$-spectral shapes.

There are also reasons to use orthogonal polynomials for $\beta$-decay energy spectra which do not apply to $\gamma$-$\gamma$ angular correlations. 
For $\gamma$-$\gamma$ angular correlations there is a natural maximum to the order of polynomial required, due to angular momentum considerations of the nuclear levels involved. 
This natural cutoff does not exist when modeling $\beta$-decay energy spectra: the order of the polynomial is arbitrary, and its choice relates to the precision of the experimental data and/or which set of theoretical corrections are applied. 
The variations in polynomial order between different experiments and theories leads to challenges comparing and interpreting the extracted coefficients.
This problem is mitigated if one uses orthogonal polynomials in $\beta$-spectra experiments and calculations; 
the value of each coefficient is ideally independent of the chosen order of the orthogonal polynomial.

A final improvement when using the orthogonal-polynomial technique is that it can provide more precise deviations from a predicted $\beta$-energy spectrum, \textit{i.e.} improved limits on measuring zero.
The crucial experimental task is to determine if there is a significant difference between the measured and expected spectra.
If one uses a power polynomial to fit an experimental spectrum, the limits on measuring zero (deviation from expectations) are often much less precise due to correlations between extracted parameters.
But extracted orthogonal-polynomial coefficients provide a more precise quantification of the deviation from expectation.


There are mathematical consistencies between the orthogonal- and power-polynomial techniques.
The total reproduced $\beta$ energy spectra fit from an orthogonal polynomial and a power polynomial have identical $\chi^{2}$ values and are mathematically equivalent for the total fit, if one is patient and tracks the correlations properly (if and when they are even reported).
The individual extracted power-polynomial coefficients can also be compared, again if one is willing and able to track the correlations between all of the power-polynomial fit coefficients.


In summary, advantages to using an orthogonal polynomial approach include: (a) extraction of a few robust coefficients which are largely independent from the maximum order of the fit; (b) reduced correlations between the extracted coefficients; (c) coefficients which have some intuitive physical meaning; (d) more robust comparisons between different theories which predict $\beta^-$ spectra; and (e) more meaningful comparison between different experimental setups that have varied detector responses.

\section{Shape Factor Extraction}

The emitted $\beta$-energy spectrum of an allowed weak decay can be modeled as \cite{hayen18}
\begin{equation} 
\begin{split} 
\label{decayEq}
	d \Gamma(W)/dW = N p W (W_{0} - W)^{2} F(Z,W) C(W) \\ \equiv B(W,Z) C(W).
\end{split}
\end{equation} 
where $p$ and $W$ are the electron momentum and total energy, $W_{0}=Q_{\beta}+m_{e}$ is the total decay energy (with speed of light, $c=1$), $m_{e}$ is the electron mass, $F(Z,W)$ is the Fermi function, $N$ is a normalization factor related to the decay rate, and $C(W)$ is the $\beta$-shape factor (or $\beta$-shape function).
$C(W)$ is commonly assumed to be of the form
\begin{equation} 
\label{usualform}
	C(W) = a_0\left(1 + a_{1} W + a_{2} W^{2} + a_{3} W^{3} + ...\right).
\end{equation}
This form of the $\beta$-shape factor will be referred to as the power polynomial expansion, or a power polynomial. 
The polynomial coefficients $a_i$ are referred to as the $\beta$-shape parameters or $\beta$-shape coefficients, and are typically what are reported in experimental measurements~\cite{mougeot15}.

\begin{figure} 
\includegraphics[width=\sizeA\linewidth]{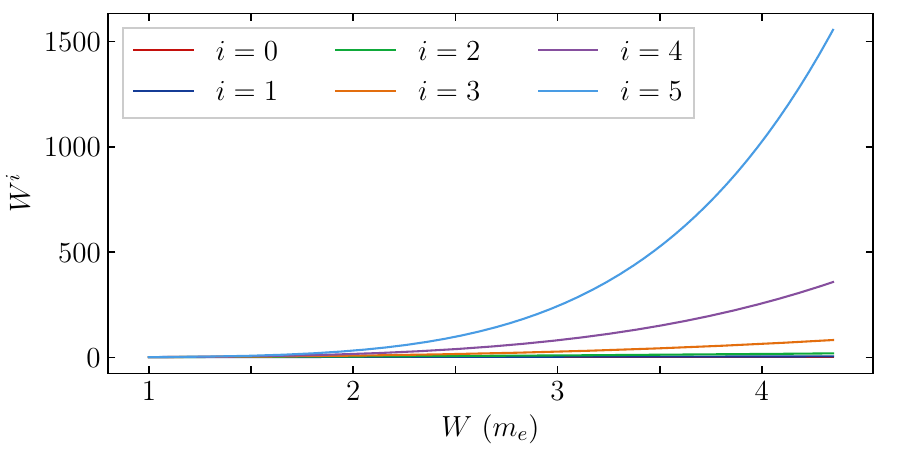}
\caption{
The first few powers, $W^{i}$, used to identify a shape factor for the  \textsuperscript{32}P $\beta$-decay energy spectrum.
\label{p32_w}
}
\end{figure}

The first few powers of $W$ for the \textsuperscript{32}P $\beta$ decay are shown in Figure \ref{p32_w}.
It is not always clear when to terminate the polynomial expansion: this varies based on statistics and what comparisons to theory are being drawn, but the appropriate termination point is properly informed by the reduced $\chi^{2}$.

Examples of techniques used to extract $C(W)$ can be found throughout the $\beta$-decay literature, where a wide variety of approaches are used to determine the $\beta$-shape parameters.
It is of interest to note that a major goal of early $\beta$-energy spectra experiments was to extract decay energies rather than the $\beta$-shape factors, though the two are connected via $\beta$-shape corrected Kurie plots.
It is not the goal of the present manuscript to fully discuss the history of $\beta$-shape parameters, but for the interested reader a wide study of reported $\beta$-shape parameters can be found in \cite{mougeot15}.

Modern examples of $\beta$-shape analysis and the extraction from experiment and theoretical model parameters can be found in \cite{kossert11, paulsen20, kostensalo23, kostensalo24, craveiro24}.
Some of the modern approaches use different polynomial parameterizations which result in the extraction of more stable coefficients, but they are still limited to the order of the polynomial considered \cite{kossert11}.

\section{Power Polynomial Example}
For illustrative purposes, experimental \textsuperscript{32}Si-\textsuperscript{32}P $\beta$-decay data is analyzed. 
\textsuperscript{32}P ($Q_{\beta}=1710.66(4)$ keV and $T_{1/2} = 14.267(6)$ days and is a single component $1^{+}$ to $0^{+}$ Gamow-Teller allowed $\beta$ decay) is produced in equilibrium with its parent \textsuperscript{32}Si ($Q_{\beta}=227.2(3)$~keV and $T_{1/2} = 157(7)$ years which is also treated as a single $0^{+}$ to $1^{+}$ Gamow-Teller allowed $\beta$ decay) \cite{chen2025}.
We note that the data shown is a set of online \textsuperscript{32}P data and does not represent a fully evaluated dataset, therefore the parameters extracted in this paper should not be considered as proper \textsuperscript{32}P $\beta$-shape parameters.
The use of experimental \textsuperscript{32}P data is to demonstrate the stability of fitting real data with the approach developed herein.

\begin{figure} 
\includegraphics[width=\sizeA\linewidth]{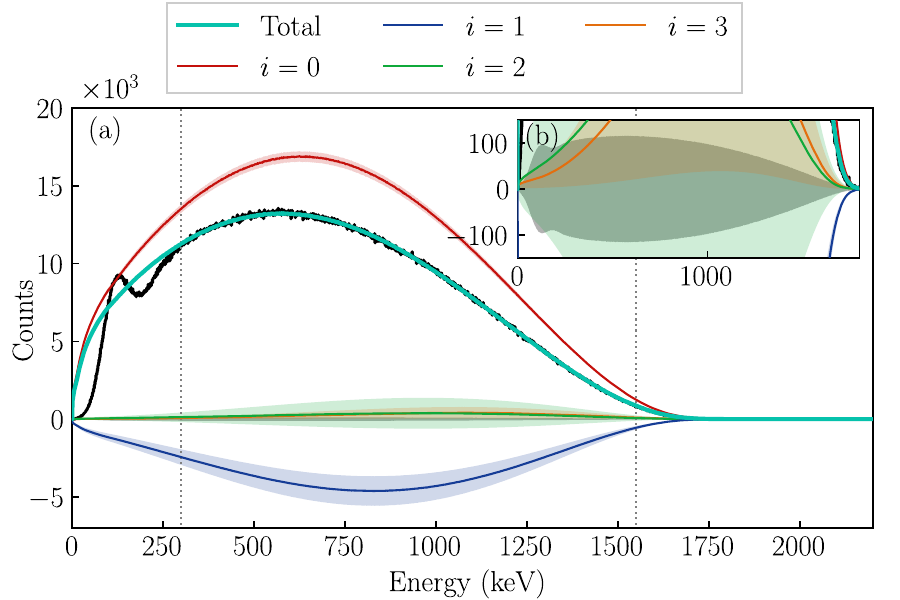}
\caption{
First few weighted power polynomials fit to \textsuperscript{32}P $\beta$ decay data. Shaded regions correspond to (uncorrelated) 1-$\sigma$ bands for each fit component. These would be what is usually reported as uncertainties in a shape-factor measurement. 
It is easy to see that there are large cancellations between the various powers of $W$.
Below  $\approx 300$~keV the parent decay and online threshold effects impact the spectrum therefore this region is not used in the fit.
The fit range is from 300 to 1550 keV.
\label{p32_powerpoly_fit}
}
\end{figure}
The result of fitting the \textsuperscript{32}P $\beta$ shape with the first four power polynomials is shown in Fig.~\ref{p32_powerpoly_fit}.
Details of the fitting technique used are provided in the Appendix.
Also provided in the Appendix are covariance and correlation matrices for various fit ranges. 
The extracted parameters as the power-polynomial order is increased are collected in Table~\ref{coeff_summary_a}.
The power-polynomial fit has large cancellations between the various fit components, especially at higher energies, which hints at large correlations between the extracted parameters. 
Also problematic is the large variation of the extracted coefficients as the maximum power of the $C(W)$ polynomial is varied. 
The large variation of each coefficient as a function of the power-polynomial order makes consistent physical interpretation challenging.

\begin{table}[h]
\sisetup{
table-alignment-mode = format,
table-number-alignment = center,
}
\caption{
\label{coeff_summary_a}
Extracted power-polynomial coefficients from online \textsuperscript{32}P data. The shape factor has the form $C(W) = a_0(1 + a_{1} W + a_{2} W^{2}  + a_{3} W^3+...) $ with each terminated at the listed maximum polynomial used in the fit. All of these terms have very large uncertainties and the physical interpretation of the coefficients is challenging as discussed in Section \ref{solving_sect}.
All results assume the speed of light is unity, $c=1$.
}
\begin{ruledtabular}
\begin{tabular}{S[table-column-width=0.8cm]|S[table-format=1.2]S[table-format=3.3] S[table-format=1.3] S[table-format=1.3] S[table-format=2.3]}
{Max} & {$a_0$} & {$a_1$} & {$a_2$} & {$a_3$} & {Red.} \\
{Poly.}  & {($\times 10^7$)} &  {$(m_e^{-1})$} & {$(m_e^{-2})$} & {$(m_e^{-3})$} & { $\chi^{2}$ } \\
\hline
{$1$} & 1.3566(4)& {-} & {-} & {-} & {25.7} \\
{$W$} & 1.6715(19) & -0.0729(3) &{-} & {-} & {1.34}\\
{$W^{2}$} & 1.828(8) & -0.134(3) & 0.0126(7) & {-} & {1.02}\\
{$W^{3}$} & 1.790(35) & -0.11(2) & 0.003(8) & 0.0012(11) & {1.02} \\
\end{tabular} 
\end{ruledtabular}
\end{table}

\section{Weighted Orthogonal Shape Functions}

To minimize the challenges facing a power-polynomial parameterization, one can change the basis used to construct the shape function $C(W)$. 
The power-polynomial approach uses the basis set $\lbrace 1, W, W^2, W^3,... \rbrace$. 
If instead one uses a basis set of functions that are orthogonal, the addition of each successive term in the expansion minimally affects the best-fit coefficients of previous terms. 

The weighted orthogonal functions $\Phi_i(W,Z)$ (or just $\Phi_i(W)$ or $\Phi_i$ in shorthand) that make up this basis are defined by 
\begin{equation}
\int_{m_{e}}^{W_{0}} B(W,Z) \Phi_{i}(W,Z) \Phi_{j}(W,Z) dW = N_{i} \delta_{ij}.
\label{orthoEq}
\end{equation}
Where $B(W,Z)$ (or just $B(W)$) is called the weight function.
If defined as in Equation \ref{decayEq}, $B(W)$ is equivalent to a pure allowed $\beta$ decay, but in general this weight function can be modeled as any allowed or forbidden $\beta$-energy spectrum including any corrections desired.

\begin{figure} 
\includegraphics[width=\sizeA\linewidth]{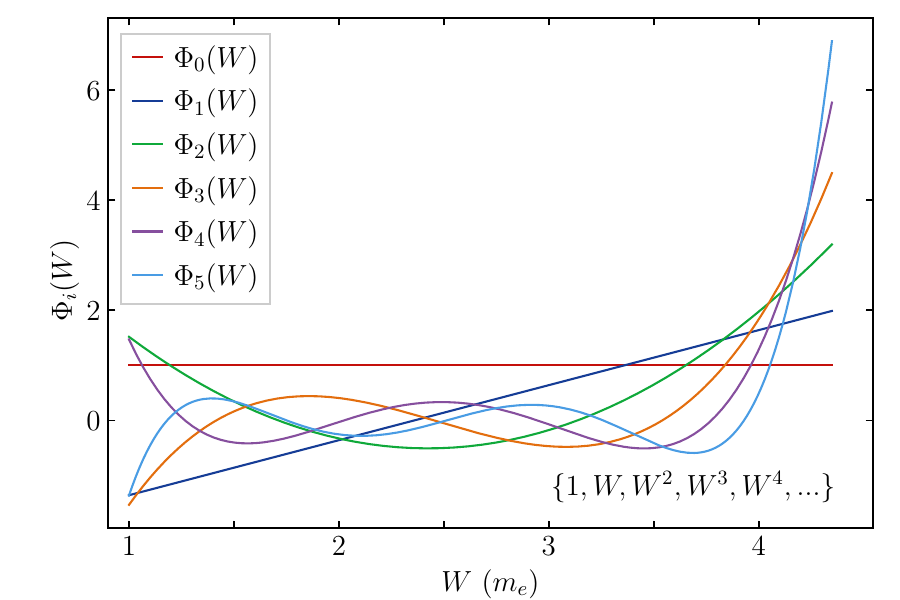}
\caption{
First few weighted orthogonal polynomials generated for studying the \textsuperscript{32}P $\beta$ decay.
The range of the $\Phi_{i}$ is between $ -0.7 \lesssim \Phi_{i}(W) \lesssim 7.0$ instead of the 2.5 order of magnitude larger range for the power polynomials shown in Figure \ref{p32_w}.
\label{p32_shapes}
}
\end{figure}

Starting with the assumption that the first polynomial $\Phi_{0}(W,Z) = 1$, the rest of the $ \Phi_{i}(W,Z)$ can be calculated using the Gram-Schmidt orthogonalization technique, using the polynomials $\left \{ 1, W, W^{2}, W^{3}, ...\right \}$ as the starting basis (the order chosen defines the resulting $\Phi_{i}$). 
Generalized discussions of weighted orthogonal polynomials are found in many intermediate linear algebra books \cite{linearalgebrabook, linearalgebrabook2}.
The resulting form of the $\beta$-shape polynomials will be referred to as an orthogonal polynomial expansion, or just as an orthogonal polynomial, and is given in Eq.~\ref{newform}
\begin{align}
\label{newform}
    C(W) = b_0\left(1 + b_1 \Phi_1(W) + b_2 \Phi_2(W) +... \right).
\end{align}
The first few of the orthogonal polynomials created using the $B(W)$ for the pure allowed $\beta$-decay of \textsuperscript{32}P are shown in Figure \ref{p32_shapes} and the first three non-constant $\Phi_{i}$ are shown in Eqs.~\ref{p32_phi1}--\ref{p32_phi3} with $W$ in electron mass units,
\begin{align}
\label{p32_phi1}
\Phi_1(W) &= -2.360 + W, \\
\Phi_2(W) &= 5.364 - 4.847W + W^2, \\
\label{p32_phi3}
\Phi_3(W) &= -12.287 + 17.161 W -7.407W^2 + W^3.
\end{align}

\noindent
A python script to generate these polynomials for  allowed decays with arbitrary decay energy and Z is provided in the supplementary material.

The stability of the extracted $b_i$ coefficients for the $\Phi_i(W)$ polynomials is shown in Table~\ref{coeff_summary_b}. The $b_i$ coefficients do not change significantly as higher-order terms are added to the expansion.
The small amount of variation present is due to the limited fitting range, which is discussed in more detail below. 
The reduced $\chi^{2}$ for the power polynomial is the same as the orthogonal polynomial reduced $\chi^{2}$ at each order if the same fitting range is used, showing the similarity of the overall fit is independent of the form of $C(W)$.

\section{Covariances, Correlations, and Fit Ranges}
The covariance over the fit range $300-1550$~keV and the correlation matrices over multiple fit ranges of $500-1550$~keV, $300-1550$~keV, and $50-1550$~keV for the power- and orthogonal-polynomial fits are shown in the Appendix. 
The different fit ranges show the fit range impact on the orthogonal-polynomial correlation matrices.
Though from an experimental perspective only the \textsuperscript{32}P component of the data should be fit, \emph{i.e.} the $300-1550$~keV fit range is preferred.
Several observations are worth discussion.

For the power-polynomial fits an increase in the variance (diagonal elements of the covariance matrix) for each coefficient is observed as the power of the polynomial increases.
The increase in variances represents the increase in the uncertainty of the extracted coefficient and if one wishes to measure a zero or small coefficient this is not optimal. 
The correlation matrix elements are all very near either $\pm1$, indicating a high linear correlation between the coefficients.
These observations about the power-polynomial coefficients are independent of the fit range.

For the orthogonal-polynomial fits the covariance matrix elements also increase as the power of the polynomial increases but are much smaller than the power-polynomial fits of the same order. 
This becomes more obvious for higher order polynomials.
The last variances, $\sigma_{j_{\mathrm{max}},j_{\mathrm{max}}}$, are always identical for both polynomial approaches. 
This is because these are the only fit components with the highest polynomial power in both cases.
For the orthogonal-polynomial fits the correlation matrix elements over the $500-1550$~keV fit range are smaller than the power-polynomial correlations, but still non-zero.
Once the range over which the orthogonal polynomials are fit becomes closer to the defined is used ($300-1550$~keV and $50-1550$~keV), the correlations drop and are nearly zero for the widest fit.
This highlights the improved robustness of the orthogonal-polynomial coefficients as well as the importance of fitting over as wide a range as possible, though a balance of using clean data must be maintained.

\section{Recognizing Statistical Limits}
\begin{figure} 
\includegraphics[width=\sizeA\linewidth]{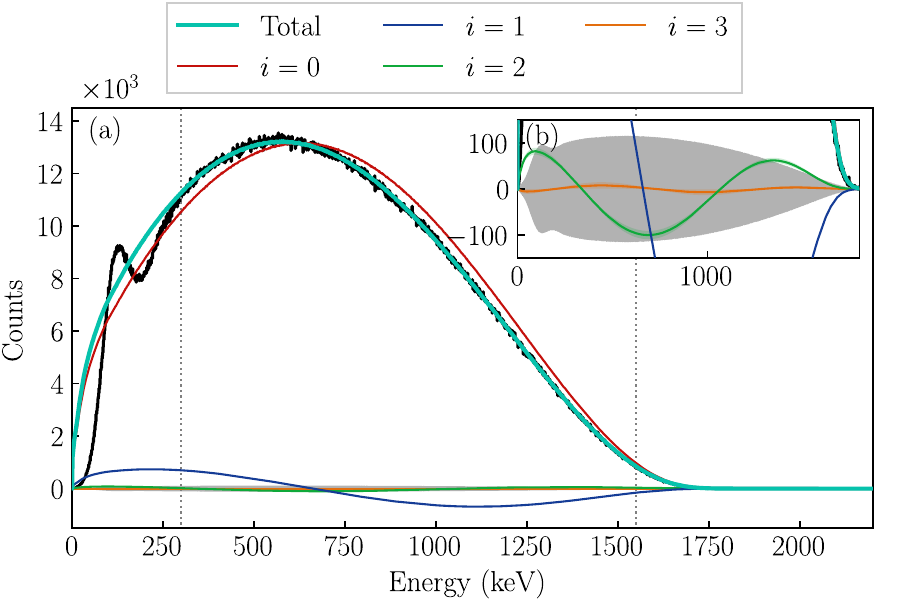}
\caption{
First few weighted orthogonal polynomials fit to online \textsuperscript{32}P $\beta$ decay data. As in Figure~\ref{p32_powerpoly_fit}, fit components have shaded regions corresponding to 1-$\sigma$ uncertainties, however in this case they are mostly narrower than the line width.
The fit range is from 300 to 1550 keV. 
Inset (b) shows that the higher order components ($i>1$) lie near the edge or well within the statistical uncertainties indicated by in the gray-shaded region.
This is the same \textsuperscript{32}P data shown in Figure \ref{p32_powerpoly_fit}.
\label{p32_orthopoly_fit}
}
\end{figure}

Figure~\ref{p32_orthopoly_fit} shows the result of fitting the \textsuperscript{32}P $\beta$-energy spectrum with the third-order orthogonal basis functions. 
The decreasing impact of higher-order terms is readily apparent in Fig.~\ref{p32_orthopoly_fit}, in contrast to the power-polynomial fit shown in Fig.~\ref{p32_powerpoly_fit} where only increased correlation is observed.
In particular, the last polynomial shown, $\Phi_{3}(W)$, is well within the statistical uncertainties of the data, shown by the gray-shaded region. 
For the current \textsuperscript{32}P data set this implies that orthogonal polynomials of the third order and beyond have overall amplitudes well within the statistical variation of the data.

\begin{table}[h]
\sisetup{
table-alignment-mode = format,
table-number-alignment = center,
}
\caption{
\label{coeff_summary_b}
Extracted orthogonal-polynomial coefficients from online \textsuperscript{32}P data for a fit range of $300-1550$~keV. 
The shape factor has the form $C(W) = b_0(1 + b_{1} \Phi_{1}+ b_{2} \Phi_{2}  + b_{3} \Phi_{3}+...)$, with each terminated at the listed maximum polynomial used in the fit.
All results assume the speed of light is in units of $c=1$.
}
\begin{ruledtabular}
\begin{tabular}{S[table-column-width=0.8cm]|S[table-format=0.3] S[table-format=3.3] S[table-format=1.3] S[table-format=1.3] S[table-format=2.3]}
{Max} & {$b_0$} & {$b_1$} & {$b_2$} & {$b_3$} & {Red.} \\
{Poly.}  & {($\times 10^7$)} &  {$(m_e^{-1})$} & {$(m_e^{-2})$} & {$(m_e^{-3})$} & { $\chi^{2}$ } \\
\hline
{$\Phi_{0}$} & 1.3566(4) & {-} & {-} & {-} & {25.3}\\
{$\Phi_{1}$} & 1.3900(4) & -0.0876(5) & {-} & {-} & {1.36}\\
{$\Phi_{2}$} & 1.3963(5) & -0.0962(6) & 0.0165(8) & {-} & {1.02} \\
{$\Phi_{3}$} & 1.3956(8) &  -0.0952(11) & 0.0153(14) &  0.0015(14) & {1.02} \\
\end{tabular} 
\end{ruledtabular}

\end{table}
\section{Moment Description of Orthogonal Polynomials}
The orthogonal basis functions $\Phi_i(W)$ can be expressed in terms of moments of the weight function $B(W,Z)$:
\begin{equation} 
\label{momentdef}
	\left<W^{n} \right> = \frac{ \displaystyle\int_{m_{e}}^{W_{0}} W^{n} B(W,Z) dW } { \displaystyle\int_{m_{e}}^{W_{0}} B(W,Z) dW} .
\end{equation}
If $\Phi_{0}(W) \equiv 1$, then 
\begin{equation} 
\label{phi1mom}
\Phi_{1}(W) = W - \left< W \right>, 
\end{equation}
and after a bit of algebra 
\begin{equation} 
\label{phi2mom}
\begin{split} 
	\Phi_{2}(W) = W^{2} - W \frac{ \left< W^{3} \right> - \left< W^{2} \right> \left< W \right> } { \left< W^{2} \right> -  \left< W \right>^{2} } \\ 
	+ \frac{ \left< W^{3} \right> \left< W \right> - \left< W^{2} \right>^{2}} {\left< W^{2} \right> - \left< W \right>^{2}} . 
\end{split} 
\end{equation}
Equation \ref{phi1mom} gives some simple physical meaning to the first two polynomials, $\Phi_{0}(W)$ and $\Phi_{1}(W)$, as the integral of the $\beta$ spectrum and the difference of the energy from the mean of the $B(W,Z)$ distribution. 
The small negative $b_1$ coefficients in Table~\ref{coeff_summary_b} indicate that the average energy of the experimental $\beta$-shape distribution is slightly lower than that of the $B(W,Z)$ weighting function. 

A simple physical interpretation of the terms in Eq.~\ref{phi2mom} is less straightforward, though we note that $\langle W^2 \rangle - \langle W \rangle^2$ is equal to the variance of the $B(W,Z)$ distribution, and that the second and third terms are related to the skewness of $B(W,Z)$.

\section{Solving for Powers of $\mathbf{W}$}
\label{solving_sect} 
One may also connect the power polynomial coefficients, $a_{i}$, with the orthogonal polynomial coefficients, $b_{i}$.
Considering only to first order (for simplicity), the two sets of parameters extracted by the two different methods by combining Eqs.~\ref{usualform} and \ref{newform} can be compared, 
\begin{align} 
\label{expansion1} \nonumber
a_0 ( 1 + a_1 W ) &= b_0(1 + b_{1} \Phi_{1})  
    \\
    \\ &= b_0 ( 1 - b_1 \langle W \rangle ) \left( 1 + 
    \frac{b_1}{1-b_1\langle W \rangle} W \right).
\end{align}
This makes it clear that the coefficients $a_0$ and $b_0$ differ depending on the order of the power polynomial chosen. At first order:
\begin{align}
\label{a0}
    a_0 &= b_0(1-b_1\langle W \rangle).
\end{align}
Also the $a_1$ and $b_1$ coefficients for a first order power polynomial are related by
\begin{align}
\label{a1}
    a_1 = \frac{b_0 b_1}{a_0} = \frac{b_1}{1-b_1\langle W \rangle},
\end{align}
which highlights how the more independent orthogonal-polynomial coefficients are connected to the correlated power-polynomial coefficients.
While $b_0$ and $b_1$ are minimally correlated (due to the orthogonal formulation of $\Phi_1$(W)), $a_0$ and $a_1$ are correlated in a non-trivial manner, linked by Equations \ref{a0} and \ref{a1}. 
The connection between the various $a_{i}$ becomes even more convoluted when using higher-order power polynomials.
This connection between the normalization and the expansion coefficient has been pointed out before in terms of isolating the Fierz term \cite{paul70} and emphasized more recently \cite{ivanov13, gonzalez2016}. 
However, when using the orthogonal-polynomial expansion the overall normalization coefficient, $b_0$, is independent of the subsequent shape coefficients $b_{i>0}$. 

\section{ Including $\mathbf{ W^{-1}}$ in the Expansion}
\begin{figure}[h]
\includegraphics[width=\sizeA\linewidth]{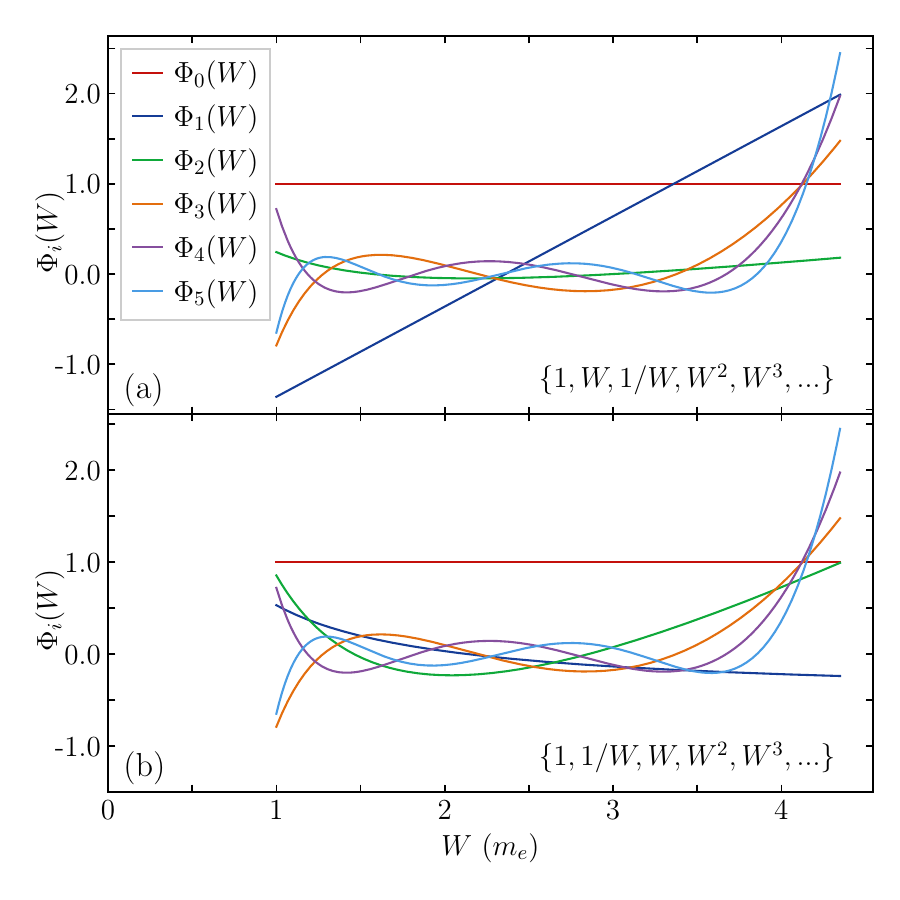}
\caption{
First few weighted orthogonal polynomials including $W^{-1}$ generated for studying the \textsuperscript{32}P $\beta$ decay. (a) Ordering of the basis functions is $\lbrace 1, W, W^{-1}, W^2...\rbrace$, i.e. the first basis function with a $W^{-1}$ term is $\Phi_2$. (b) Ordering of the basis functions is $\lbrace 1, W^{-1}, W, W^2...\rbrace$, i.e. the first basis function with a $W^{-1}$ term is $\Phi_1$.
\label{p32_shapes_m1}
}
\end{figure}

Considering the importance of the Fierz term, how to add a $W^{-1}$ term to the expansion should be commented on.
Adding a $W^{-1}$ term merely creates a different set of orthogonal functions than the orthogonal polynomials considered above. 
Rather than starting the Gram-Schmidt process with the basis $\lbrace 1, W, W^2, W^3... \rbrace$, a different set $\lbrace 1, W, W^{-1}, W^2, W^3,... \rbrace$ may be chosen. 
The resulting $\Phi_i$ functions are orthogonal, and are shown in Fig.~\ref{p32_shapes_m1}(a). 
The inclusion of such a term may be useful to extract improved limits on the Fierz term~\cite{paul70,gonzalez2016}. 
The choice of where to add the $W^{-1}$ term is not unique, it is also possible to choose a different ordering of the starting functions such as $\lbrace 1, W^{-1}, W, W^2, W^3,... \rbrace$. 
This results in different orthogonal functions $\Phi_i$, which are shown in Fig.~\ref{p32_shapes_m1}(b). 
A more detailed discussion of the $W^{-1}$ component is not the main focus of this manuscript, but can be accommodated with extension of the technique offered herein. 

\section{Weight Function Choice}

Up to this point in the manuscript, in order to demonstrate the usefulness of the orthogonal-polynomial approach, a pure allowed $\beta$ shape with no corrections is chosen for the weight function, $B(W)$. 
This should be a reasonable approximation to the \textsuperscript{32}P $\beta$ decay shape.
However, since the extracted higher order orthogonal-polynomial coefficients are statistically non-zero, ($b_{i>0} \neq 0$), this implies that a pure allowed spectrum does not precisely describe our preliminary experimental $\beta$ spectrum. 
This is due to missing corrections to the allowed $\beta$ shape \cite{hayen18} and also effects such as detector Bremsstrahlung and electron back-scattering which may not be fully accounted for in the response function.
Once all of these corrections are identified, they can be included directly in the $B(W)$ weight function, at that point all correction coefficients should be nearly consistent with zero. 
However, in the case where all such corrections are applied and the correction coefficients are not consistent with zero, the orthogonal-polynomial method has an improved ability to identify and quantitatively describe the differences between the measured and expected spectra.

As an example suppose that the complete measured shape is predicted (\textit{i.e.} assumed) to be 
\begin{equation} 
\label{differentWeight}
    B'(W) = N p W (W_{0} - W)^{2} F(Z,W) C_2(W),
\end{equation}
where $C_2(W)$ is the orthogonal polynomial defined with the fit parameters in Table II.
The orthogonal polynomials can be generated for this weight function, $\Phi^{'}_{i}(W)$. 
The resulting second order shape functions extracted are
\begin{equation} 
\label{differentWeightParametersOP}
    C'(W) = 1.0 - 0.00027(66) \Phi^{'}_{1} - 0.0007(9) \Phi^{'}_{2},
\end{equation}
for the orthogonal polynomial and for the power polynomial the result is
\begin{equation} 
\label{differentWeightParametersPP}
    C'(W) = 1.0 + 0.0029(46) W - 0.0007(9) W^{2}.
\end{equation}
In addition to the stability of the $b_{0}$ coefficient, the uncertainty of the first order orthogonal polynomial coefficient is over a factor of 10 smaller than the limit on the first order power polynomial parameter while the second order has the same coefficient and uncertainty for both shape factors.
This reduction is due to the improved orthogonality of the $\Phi_{i}$ compared to the power polynomials.

Fundamentally any choice of weight function can be used, though choosing a physically relevant one is likely desirable.
For a first-forbidden unique decay it would be appropriate to use a pure first-forbidden unique shape factor, with whatever additional corrections are expected.
For first-forbidden non-unique decays a choice that includes the nuclear matrix element influence would be appropriate.
The accuracy of corrections to the predicted $\beta$-shape can be tested by including them in the $B(W)$ function and comparing the magnitude of the extracted $b_{i>0}$ coefficients and whether they are consistent with zero or not.
If the coefficients are not consistent with zero then the assumed weight function needs modification, though the source of the appropriate correction is only hinted at, as there may be more than one type of correction that behaves as the identified nonzero terms.




\section{Conclusion}
We offer an alternate orthogonal-polynomial expansion for the $\beta$-decay shape function, $C(W)$, instead of the usual power-polynomial expansion. 
The orthogonal-polynomial expansion has several helpful features when compared to the power-series expansion of $C(W)$. The orthogonal-polynomial approach
\begin{itemize}
\item Minimizes the dependence of the coefficient values on the order of the expansion, allowing for easier comparison between different experiments and theories.
\item Reduces of the covariances and correlations between the extracted coefficients.
This reduction is improved if more of the full orthogonal polynomial fit range is used. 
\item Offers physical meaning, with respect to the weight function, $B(W,Z)$, for each coefficient independent of the polynomial order used.
\item Offers a qualitative handle on when the expansion is fitting statistical variations.
\end{itemize}
In short, the extracted orthogonal-polynomial coefficients are more robust than other shape-function parameterizations.
This type of analysis approach may have applicability for extracting consistent shape factors for low resolution measurements such as reactor antineutrino spectra detected via inverse-$\beta$ decay.

\section{Acknowledgments}
This work is supported by the DOE Office of Science and by the Nuclear Data Inter-Agency Working Group (NDIAWG) Funding Opportunity-2440.
This material is based upon work supported in part by the U.S. Department of Energy, Office of Science, Office of Nuclear Physics under Contracts 
 No. DE-AC05-00OR22725 (ORNL) and No. DE-FG02-96ER40983 (UTK).

\bibliography{DecayChain}
\clearpage
\appendix

\section{Experimentally Extracting $\mathbf{\beta}$-Shape Coefficients}
The following describes the method used to extract shape values from an experimentally measured $\beta$-energy spectra, $D(E)$,
\label{deconv}
\begin{align} \nonumber
D(E) &= \int_{m_{e}}^{W_{0}} R(E,W)  p W (W_{0} - W)^{2} \times \\ &\qquad \qquad \qquad F(Z,W) C(W) dW,
\end{align}
where $m_{e}$ is the electron mass, $W_{0} = Q_{\beta} + m_{e}$ is the total $\beta$-decay energy, $R(E,W)$ is the detector response function, $p$ and $W$ are respectively the electron momentum and total energy, $F(Z,W)$ is the Fermi function, and $C(W)$ is the $\beta$-shape function.
The detector response function maps the emitted total electron energy, $W$ (in electron mass units), to the energy detected in the detector, $E$ (in keV).
The $\beta$-shape function can be written as 
\begin{equation} 
\label{bseform}
C(W) = \sum_{i} c_{i} X_{i}(W),
\end{equation} 
where the $X_{i}(W)$ are polynomials that can be chosen as a straight power of total electron energy, $W^{i}$, or as an orthogonal polynomial, $\Phi_{i}(W)$.
This leads to 
\begin{equation} 
\begin{split}
\label{deconv1}
D(E) = \sum_{i} c_{i} \int_{m_{e}}^{W_{0}} R(E,W) p W (W_{0} - W)^{2} \\ \times F(Z,W) X_{i}(W) dW,
\end{split}
\end{equation} 
where the integrals can be performed numerically and the $c_{i}$ can be solved for by minimizing the $\chi^2$.

Whether the response function, $R(E,W)$, should be included in the weight function, $B(W)$, is not obvious.
Doing so would force an experimental form factor onto the $\Phi_i$ basis.
This makes the extracted shape coefficients less general and more difficult to interpret without the experimental connection. 
The slightly different $\Phi_{i}$ would be orthogonal and hence the extracted coefficients, $b_{i}$ would be more independent from each other. 
To minimize experimental influence on the extracted coefficients, we have chosen to not include the detector response in the weight function, $B(W)$. 
This with the orthogonal basis defined over the full range $W \in [m_e, W_{0}]$, while the experimental data was only fit over $E \in [300,1550]$ keV means that the $\Phi_i$ functions are \textit{not quite} orthogonal with respect to the fits conducted. 
Consequently, the coefficients given in Table~\ref{coeff_summary_b} change slightly as the maximum order is increased. 
Nonetheless, the robustness of the results presented in this manuscript are hopefully convincing and indicate that not including the response function is a valid choice, at least for our experimental setup and response function.
Further investigation into the impact of the detector response function is warranted.

If one considers comparison with theory, then the response function can be taken as a delta function, $\delta( W - E )$ and Equation \ref{deconv1} reduces to
\begin{equation} 
\begin{split}
\label{deconv1theory}
D(E) = \sum_{i} c_{i} p E (W_{0} - E)^{2} F(Z,E) X_{i}(E),
\end{split}
\end{equation} 
where E represents the electron total energy.
Using a delta function as the response function identifies the detected total electron energy as the same as the emitted total electron energy, which experimentally are not the same.

\section{Covariance Matrices}

The covariance matrices for both power and orthogonal polynomial orders 1 through 4 are shown below.
Each covariance matrix is associated with the fit parameters $c_{i}$ in Equation \ref{deconv1} and are related to the $a_{i}$ and $b_{i}$ in Tables \ref{coeff_summary_a} and \ref{coeff_summary_b} by $a_{0} = c_{0}$ along with $a_{i} = c_{i} / c_{0}$ for $i \neq 0$ for the power polynomials and similarly for the orthogonal polynomials $b_{0} = c_{0}$ with $b_{i} = c_{i} / c_{0}$ for $i \neq 0$. 
The coefficients are solved for using weighted least squares over the fit range $300-1550$~keV.
All $c_{i}$ and covariance matrix elements are in units of $10^{7}$. For example, the power polynomial coefficient  $c_{3}$ is reported as $c_{3} = 0.0021\times 10^7 \pm \sqrt{36 \times 10^{7} } = 2.1(19) \times 10^{4}$.

\paragraph{Power Polynomial Covariance Matrices}
\begin{align}
c_{0} &= ( 1.3566 ),  \sigma^{2}_{0,0} = ( 1.65 ) \\
\begin{pmatrix} c_{0} \\ c_{1} \end{pmatrix} &=  
\begin{pmatrix} 1.6715 \\ -0.1218 \end{pmatrix}, 
\begin{pmatrix} 
34 & -13 \\ 
-13 &  4.9 
\end{pmatrix} \\
\begin{pmatrix} c_{0} \\ c_{1} \\ c_{2} \end{pmatrix} &=  \begin{pmatrix} 
1.8276 \\ -0.2449 \\ 0.0231 
\end{pmatrix}, 
\begin{pmatrix}  
638 & -488 & 89 \\ 
-488 & 380 & -70 \\ 
89 & -70 & 13 
\end{pmatrix} \\
\begin{pmatrix} c_{0} \\ c_{1} \\ c_{2} \\ c_{3} \end{pmatrix} &= 
\begin{pmatrix} 
1.7900 \\ -0.20 \\ 0.0059 \\ 0.0021 
\end{pmatrix}, 
\begin{pmatrix}  
12260 & -14300 & 5360 & -648 \\ 
-14300 & 16800 & -6332 & 770 \\ 
 5360 &  -6332 & 2400 & -294 \\ 
-648 & 770 & -294 & 36.1 \\ 
\end{pmatrix}
\end{align}

\paragraph{Orthogonal Polynomial Covariance Matrices}
\begin{align}
c_{0} &= ( 1.3566 ),  \sigma^{2}_{0,0} = ( 1.65 ) \\
\begin{pmatrix} c_{0} \\ c_{1} \end{pmatrix} &=  
\begin{pmatrix} 1.3900 \\ -0.1218 \end{pmatrix}, 
\begin{pmatrix} 
2.02 & -1.33 \\ 
-1.33 & 4.9 \end{pmatrix} \\
\begin{pmatrix} c_{0} \\ c_{1} \\ c_{2} \end{pmatrix} &=  
\begin{pmatrix} 1.3963 \\ -0.1344 \\ 0.0231 \end{pmatrix}, 
\begin{pmatrix} 
3.00 & -3.30 & 3.61 \\ 
-3.30 & 8.79 & -7.19 \\ 
3.61 & -7.19 &  13.2 
\end{pmatrix} \\
\begin{pmatrix} c_{0} \\ c_{1} \\ c_{2} \\ c_{3} \end{pmatrix} &=  \begin{pmatrix} 1.3956\\ -0.132 \\ 0.021 \\ 0.0021 \end{pmatrix}, 
\begin{pmatrix}  
6.33 & -10.7 & 12.2 & -11.0 \\ 
-10.7 & 25.2 & -26.3 & 24.4 \\ 
 12.2 & -26.3 & 35.5 & -28.4 \\ 
-11.0 & 24.4 & -28.4 & 36.1 \\ 
\end{pmatrix}
\end{align}

The $c_{j_{\mathrm{max}}}$ and each covariance matrix element in the lower right-hand corner, $\sigma_{j_{\mathrm{max}},j_{\mathrm{max}}}$, in each corresponding covariance matrix for both the power and the orthogonal polynomial are always identical which is a consequence of the orthogonal-polynomial normalization chosen.
This identity is not necessarily true for the $a_{i}$ and $b_{i}$.

\section{Correlation Matrices}

The correlation matrix elements are defined as
\begin{equation} 
\label{corrdef}
	\rho^{(k)}_{ij} = \frac{ \sigma_{ij} }{ \sqrt {\sigma_{ii} \sigma_{jj}} }, 
\end{equation}
where $(k)$ refers to the maximum order of the fit. The various correlation matrices for linear through cubic polynomial orders are shown below. 
The covariance matrices are from using a weighted least squares over three different ranges, $500-1550$~keV, $300-1550$~keV, and $50-1550$~keV.
The different fit ranges show the reduction of the orthogonal-polynomial correlations as the range approaches the full energy range the orthogonal polynomials are defined over.
All numbers are rounded to two decimal places for readability.

The following are the correlation matrices for the $c_{i}$ over the fit range $500-1550$~keV.
\paragraph{Power Polynomial Correlation Matrices}
\begin{align}
\rho^{(2)} &= \begin{pmatrix}  
1 & -0.98 \\ 
-0.98 &  1 
\end{pmatrix} \\
\rho^{(3)} &=\begin{pmatrix}  
1 & -1.00 & 0.98 \\
-1.00 & 1 & -1.00 \\
 0.98 & -1.00 & 1 \\
\end{pmatrix} \\
\rho^{(4)} &= \begin{pmatrix} 
1 & -1.00 & 0.99 & -0.98 \\
-1.00 & 1 & -1.00 & 0.99 \\
 0.99 & -1.00 & 1 & -1.00 \\
-0.98 & 0.99 & -1.00 & 1 
\end{pmatrix}
\end{align}

\paragraph{Orthogonal Polynomial Correlation Matrices}

\begin{align}
\rho^{(2)} &= \begin{pmatrix} 
1 & -0.64 \\ 
-0.64 & 1 
\end{pmatrix}\\
\rho^{(3)} &= \begin{pmatrix} 
1 & -0.90 & 0.84 \\
-0.90 & 1 & -0.87 \\
0.84 & -0.87 & 1 \\
\end{pmatrix} \\
\rho^{(4)} &= \begin{pmatrix}  
1 & -0.98 & 0.97 & -0.91 \\
-0.98 & 1 & -0.98 & 0.93 \\
 0.97 & -0.98 & 1 & -0.93 \\
-0.91 & 0.93 & -0.93 & 1 
\end{pmatrix}
\end{align}

The following are the correlation matrices for the $c_{i}$ over the fit range $300-1550$~keV.
\paragraph{Power Polynomial Correlation Matrices}
\begin{align}
\rho^{(2)} &= \begin{pmatrix}  
1 & -0.98 \\ 
-0.98 &  1 
\end{pmatrix} \\
\rho^{(3)} &=\begin{pmatrix}  
1 & -0.99 & 0.97 \\
-0.99 & 1 & -0.99 \\
 0.97 & -0.99 & 1 \\
\end{pmatrix} \\
\rho^{(4)} &= \begin{pmatrix} 
1 & -1.00 & 0.99 & -0.97 \\
-1.00 & 1 & -1.00 & 0.99 \\
 0.99 & -1.00 & 1 & -1.00 \\
-0.97 & 0.99 & -1.00 & 1 
\end{pmatrix}
\end{align}

\paragraph{Orthogonal Polynomial Correlation Matrices}
\begin{align}
\rho^{(2)} &= \begin{pmatrix} 
1 & -0.37 \\ 
-0.37 & 1 
\end{pmatrix}\\
\rho^{(3)} &= \begin{pmatrix} 
1 & -0.58 & 0.55 \\
-0.58 & 1 & -0.63 \\
0.55 & -0.63 & 1 \\
\end{pmatrix} \\
\rho^{(4)} &= \begin{pmatrix}  
1 & -0.81 & 0.79 & -0.71 \\
-0.81 & 1 & -0.86 & 0.80 \\
 0.79 & -0.86 & 1 & -0.77 \\
-0.71 & 0.80 & -0.77 & 1 
\end{pmatrix}
\end{align}

The following are the correlation matrices for the $c_{i}$ over the fit range $50-1550$~keV.
\paragraph{Power Polynomial Correlation Matrices}

\begin{align}
\rho^{(2)} &= \begin{pmatrix}  1 & -0.96 \\ -0.96 &  1 \end{pmatrix} \\
\rho^{(3)} &=\begin{pmatrix}  
1 & -0.98 & 0.95 \\
-0.98 & 1 & -0.99 \\
0.95 & -0.99 & 1 \\
\end{pmatrix} \\
\rho^{(4)} &=\begin{pmatrix} 
1 & -0.99 & 0.97 & -0.95 \\
-0.99 & 1 & -0.99 & 0.98 \\
 0.97 & -0.99 & 1 & -1.00 \\
-0.95 & 0.98 & -1.00 & 1 
\end{pmatrix} 
\end{align}

\paragraph{Orthogonal Polynomial Correlation Matrices}
\begin{align}
\rho^{(2)} &= \begin{pmatrix} 
1 & -0.07 \\ 
-0.07 & 1 
\end{pmatrix}\\
\rho^{(3)} &= \begin{pmatrix} 
1 & -0.07 & 0.05 \\
-0.07 & 1 & -0.06 \\
0.05 & -0.06 & 1 \\
\end{pmatrix} \\
\rho^{(4)} &= \begin{pmatrix}  
1 & -0.07 & 0.05 & -0.01 \\
-0.07 & 1 & -0.07 & 0.16 \\
 0.05 & -0.07 & 1 & -0.04 \\
-0.01 & 0.16  & -0.04 & 1 
\end{pmatrix}
\end{align}

The power-polynomial coefficients are always highly correlated, independent of the fit range chosen.
The orthogonal-polynomial fit parameters over the reduced fit ranges show some correlation, though they have smaller covariances and correlations than the same fit range power-polynomial matrices.
This leftover correlation is predominantly due to the fit range not matching the range over which the the orthogonal polynomials are defined ($W \in [m_e,W_{0}]$).
When a fit range that is much closer to the orthogonal polynomial range is used, the orthogonal-polynomial correlation matrix elements are greatly reduced, whereas the power-polynomial correlation matrix elements remain near $\pm 1$.
This reduction in the off-diagonal correlation elements varies smoothly as the range of the fit approaches the full range of the orthogonal polynomial definition.
Though this is constrained by the valid data fit range, i.e. the need to avoid \textsuperscript{32}Si parent contamination below $250$ keV and to avoid detector nonlinear affects below $300$ keV. 

\begin{center}

\end{center}

\newpage
This manuscript has been authored by UT-Battelle, LLC under Contract No. DE-AC05-00OR22725 with the U.S. Department of Energy. The United States Government retains and the publisher, by accepting the article for publication, acknowledges that the United States Government retains a non-exclusive, paid-up, irrevocable, worldwide license to publish or reproduce the published form of this manuscript, or allow others to do so, for United States Government purposes. The Department of Energy will provide public access to these results of federally sponsored research in accordance with the DOE Public Access Plan (http://energy.gov/downloads/doe-public-access-plan).


\end{document}